LONG-TERM EFFECTS OF FEMALE TEACHER ON HER PUPILS' SMOKING BEHAVIOUR LATER IN LIFE

Eiji Yamamura*

*Department of Economics, Seinan Gakuin University, 6-2-92 Nishijin, Sawara-ku, Fukuoka 814-8511, Japan*

The author confirms that there are no conflicts of interest to declare.

*Correspondence to the author at the Department of Economics, Seinan Gakuin University, 6-2-92 Nishijin, Sawara-ku, Fukuoka 814-8511, Japan

Tel:+81-(0)92-823-4543

Fax: +81-(0)92-823-2506

E-mail: yamaei@seinan-gu.ac.jp.

## Summary

In Japan, teacher and student is randomly matched in the first year of elementary school. Under the quasi-natural experimental setting, we examine how learning in female teacher homeroom class in the elementary school influence pupils' smoking behavior after they become adult. We found that pupils are unlikely to smoke later in life if they belonged to female teacher homeroom class in pupil's first year of school.

# 1. INTRODUCTION

Various researchers found that an individual's smoking behavior is influenced by circumstances and norms shared by people (Yamamura 2011; Jusot et al. 2013; Rodríguez-Planas and Sanz-de-Galdeano 2019). For instance, parents have a significant effect on children's smoking behavior (e.g., Göhlmann et al. 2010; Loureiro et al. 2010; Darden and Gilleskie 2016; Yamamura 2020). Children mimic their parents' behavior and attitude toward smoking[1]. Meanwhile, gender identity leads to differences in behavior (Akerlof and Kranton 2000). The transmission mechanism of smoking differs according to parent-child gender matches (e.g., Loureiro et al. 2010; Darden and Gilleskie 2016; Yamamura 2020). The mechanism possibly exists not only in parents-child relation but also teacher-student relation. Especially, early childhood education has a long-term impact on pupil's life course (Heckman et al., 2013). However, no research investigated it. Accordingly, the present paper explores how teacher's

---

[1] Parents' values were transmitted to the his or her own children (Albanese et al 2016).

gender in the elementary school influenced pupils' smoking behavior later in life.

For this purpose, we independently conducted internet-survey to gather data of respondents' smoking behavior and their teachers' genders of the home-room teacher when they were elementary school pupils. In Japan, home-room teacher teach pupils all subjects and eat lunch together with pupils. That is, pupils spend most time with home-room teacher in the life of elementary school. Existing works reported that female cigarette consumption is smaller than male (Yamamrua and Tsutsui 2019, Yamamura 2020). Hence, female adults are expected to leads children to reduce consumption of cigarettes when they become adult. In Japan, pupils are randomly assigned to home-room class when they entered the school. Hence, genders of home-room teachers are randomly and exogenously assigned in the first grade, although, in higher grades, teachers and pupils were matched using information of pupils from past grades. Under the quasi-experimental setting, I analyzed the long-term effect of gender of the teacher in the first grade on pupils' smoking behavior after they become adult. Using the independently collected data, I conducted regression estimations. Key findings are: (1) consumption of cigarettes was lower for pupils who had a female home-room teacher in the first year of the elementary school. We divided the effect into non-smoker and reducing the amount of consumption. The pupils of female teacher are more likely to non-smoker. However, their consumption is equivalent to others if they smoked.

## 2. METHODS

## 2.1 Data

I conducted an internet survey in October 2016. The Nikkei Research Company was commissioned to conduct it. I conducted an internet survey wherein I asked 12,176 adults aged 20-65 about their smoking behavior, and then I obtain 10,710 observations which provided information about smoking behavior. Further, we also asked about genders of their home-room teacher in the elementary school, and around half of them did not reply. Hence, observations reduced to 5662. In addition, respondents are requested to answer various questions such as educational background, household income, current residential prefectures, residential prefectures when they were 6 years old, marital status, job status and structure of families. In addition, I collected the variables to capture educational curriculum such as working and learning together in the elementary school because the teaching practice influenced pupils (Algan et al, 2013). According to Official Report on School Basic Survey[2], 99 % of pupils enter the public elementary school in Japan. In the dataset, all respondents entered the elementary school.

Consequently, 4,775 observations were gathered to provide data for this study. Therefore, attention should be paid to the possibility of selection bias.

## 2.2 Model

It is critical whether pupils are randomly assigned to class regardless of teacher's characteristics.

---

[2]Official cite of Ministry of Educations, Cultural, Sports, Science and Technology, JAPAN. https://www.mext.go.jp/a_menu/koutou/shinkou/main5_a3.htm. Accessed on Jan 11, 2021.

At least, in the first grade of public school of Japan, parents cannot select the class by considering characteristics of teacher[3]. Meanwhile, teacher cannot also select child because they do not have information about characteristics of child. From viewpoint of school, teachers know about pupils' characters and dispositions by teaching them and observing their behavior during the school life. In the case there is trouble between them, next year pupils are assigned to presumably more suitable teacher. If a pupil was inappropriately matched with a female teacher in the past, he/she might be assigned to a male teacher class in next year. Therefore, the assignment to a women's class is random and exogenous only in the first grade (Yamamura and Tsutsui, 2019).

The probability of being assigned to female teacher class possibly depends on female teacher rate in the elementary school in the area where respondents resided at the year of entering school, even if pupils are randomly assigned to teacher when they entered the school. Using official surveys, we collected the number of both male teachers and female teachers for 47 prefectures in different years, which enables us to calculate female teacher rate[4]. In the questionnaire, we asked respondent's residential prefecture at the 6 years old. We then matched the ratio of female teachers with the respondents by considering their years of entering school and their respective prefecture when pupils entered the elementary school[5]. Table 1 shows that female teacher rates ranged between 0.24 and 0.73, indicating wide variation of probability being assigned to female teacher class according to time and place.

---

[3] The exception is case of children with mental retardation.
[4] "Report on School Basic Survey (various years)"
[5] Those who entered elementary school in 1963 and in the Kyoto prefecture. We matched the ratio of female teachers in Kyoto, and 1965 is the year closest to 1963 in our sample.

To assess the long-term effects of female teacher on pupils' smoking behavior later in life, the estimated function takes the following form:

$$Smoke_i = \alpha_0 + \alpha_1 \text{ Female teacher in 1st year}_i + \alpha_2 \text{ Years of female teacher in 2-6}^{th}{}_i + X'_i B + u_i.$$

where *Smoke $_i$* represents the dependent variables for individuals, and *i and α* represent the marginal effect of independent variables. Both an upper limit (41 cigarettes per day) and a lower limit (0 cigarettes per day) have been included. Therefore, we used the two-limit Tobit model for estimation. Various control variables have been included and expressed as vector X.

The key independent variable is *Female teacher in 1st year* because this captures effects of female teacher which was randomly assigned to pupil's class. Its coefficient is predicted to have a negative sign if female teacher leads pupil not to smoke. To control influence of female teacher class in higher grades, we included *Years of female teacher in 2-6$^{th}{}_i$* which aggregate years in higher grades during the period of elementary school. In alternative specification, instead of *Years of female teacher in 2-6$^{th}{}_i$*, we simply added 5 dummies of female teacher class in higher grades.

Vector of the control variables is denoted by $X_i$ and the vector of the estimated coefficients is denoted by B. $X_i$ included female teacher rates of respondent's residential prefecture when respondent entered the elementary. Female dummy is included to control respondent's gender difference of smoking behavior. From existing works (Yamamrua and Tsutsui 2019, Yamamura 2020), this is predicted to have negative sign.

Having a child reduced incentive to smoke (Yamamura and Tsutsui, 2019) and so number children was included as control variables. It is plausible that family conditions in childhood formed preferences and influenced behavior later in life (e.g., Washington, E., 2008; Oswald and Powdthavee 2010; Borrel-Porta et al., 2019). Habit of childhood in the family is captured by frequencies of eating breakfast with family. Habit of eating breakfast help child to keep rhythm of life, forming healthy life style. Therefore, regularly eating breakfast in the childhood causes child not to smoke later in life. Further, the number of brothers and that of sisters are include.

We also controlled group work and pro-competition curricula because specific educational features such as teaching practices influenced pupils' preferences even after they became adult (e.g., Milligan et al., 2004; Aspachs-Bracons et al., 2008; Algan et al. 2013). We disentangled teacher's gender of home-room class from the practices in class. In addition, control variables are 7 dummies for educational background, ages, square of ages, household income dummies, marital status dummies, current residential prefecture dummies.

For closer examination, number of cigarettes that an individual consumed were analyzed in a two-stages decision-making process (Yamamura and Tsutsui 2019; Yamamura 2020). In the first stage, the individual determines whether he/she smoke or not. In the second stage, the individual decides how many cigarettes he/she will consume. Therefore, as an alternative model, we employed a Heckman's two-stages model. In the first stage, the Probit model was used because the dependent variable is the binary dummy variable. After selection, in the second stage, the Ordinary Least Square (OLS) model should be used. In the first stage, the set of independent variables is equivalent to that of the Tobit model. In the second stage, number of children are excluded from the set of independent variables in the OLS estimation because number of children is not related to the quantity of cigarette consumption in existing research (Yamamura and Tsutsui 2019). In the second stage, as dependent variable, we used the log-formed dependent variable for convenience of interpretation because the dependent variable does not have a 0 value.

## 3. RESULTS

Table 1 displays the basic statistics and definitions of key variables in this paper. Key variable is *Smoke* which shows wide variation of cigarette consumption. However, mean of *Smoke dummy* is 0.16, implying that only 16 % is smoker. In other words, 84 % is non-smoker and so it is critical whether one smoke or not.   Table 2 indicates that pupils who belong to the female teacher class is less likely to smoke than male teacher lass in the first grade. However, surprisingly, there is no difference of cigarette consumption in higher grades.

We report regression results in Tables 3 and 4. Various control variables are included but most of them do not show statistical significance. Hence, we only report results of key variables. Table 3 reports results of two-limits Tobit estimation. In columns (1) to (3), *Female teacher in first year* show the negative sign and statistical significance at the 1 % level. However, other variables to capture female teacher class in higher grades are not statistically significant at all. This reflects that earlier childhood has greater influence on smoking behavior later in life. Absolute values of *Female teacher in first year* are around 3.50. Hence, cigarette consumption of pupils of the female teacher class is smaller by 3.5 cigarettes per day. As for control variables, consistent with prediction, *Breakfast childhood, number of children* and *Female* have the significant negative sign.

Turning to Table 4, we see results of Heckman model. In the second stage, results of variables do not show statistical significance. Therefore, consumption of cigarettes can be explained by whether one is smoker or non-smoker. Hence, we discuss the results of the first stage.

In line with Table 3, we observe the negative sign of *Female teacher in first year* and statistical significance at the 1 % level. However, with exception of the first grade, female teacher class did not influence probability of smoking at all. Marginal effect of *Female teacher in first year* are around 0.04 or 0.05. Hence, pupils of the female teacher class in the first grade is less likely to be smoker by 4 or 5 %. Concerning control variables, we observe the significant negative sign of *Breakfast childhood, number of children* and *Female*. Marginal effect of *Breakfast childhood* is 0.007, implying that an increase of eating breakfast in a week during the elementary school period reduces 0.7 % probability of smoking. Hence, in compared with person never eating breakfast, eating breakfast everyday reduces about 5% probability of smoking. This is almost equivalent to effect of belonging to the female teacher class in the first grade. Marginal effect of *number of children* is 0.01, implying that having a child reduces 1 % reduces probability of smoking. Marginal effect of *Female* is 0.13, suggesting that female is less likely to smoke by 13 % than male. Considering *Female teacher in first year* and *Female* together indicates that effect of belonging to the female teacher class in the first grade is approximately one-third of that of person's gender difference. This implies that transmission form female teacher is sizable in the early childhood education.

# 4. CONCLUSION AND DISCUSSION

Based on independently collected data, we examined the influence of female home-room teacher on her pupil's smoking behavior after being adult under the quasi-natural experiment where teachers were randomly assigned to pupils when they entered the elementary school. Key findings are that pupils are unlikely to smoke later in life if they belonged to female teacher homeroom class in pupil's first year of school.

The contribution of the present paper is to bridge education and health economics to provide new evidence that female teacher improved her pupils' health later in life through preference formation to reduce probability of pupils' smoking later in life. This is consistent with existing works made it evident that surrounding adults has long-term impact on child's smoking behavior later in life (Darden & Gilleskie 2016; Loureiro et al., 2010; 2020).

Table 1. Definitions of key variables and their basic statistics

| Variables | Definition | Mean | Standard deviation | Min. | Max. |
|---|---|---|---|---|---|
| *Dependent Variables* | | | | | |
| *Smoke* | Number of cigarettes that the respondent smoked per day. From 0 (Not at all) to 41 (Equal to or more than 41 cigarettes). | 1.86 | 5.51 | 0 | 45 |
| *Smoke dummy* | Equals 1 if you smoke, 0 otherwise | 0.16 | 0.37 | 0 | 1 |
| *Female teacher in first year.* | Equals 1 if class teacher is female at the first grade in elementary school, 0 otherwise | 0.81 | 0.39 | 0 | 1 |
| *Female teacher in second year.* | Equals 1 if class teacher is female at the second grade in elementary school, 0 otherwise | 0.73 | 0.44 | 0 | 1 |
| *Female teacher in third year.* | Equals 1 if class teacher is female at the third grade in elementary school, 0 otherwise | 0.58 | 0.49 | 0 | 1 |
| *Female teacher in fourth year.* | Equals 1 if class teacher is female at the fourth grade in elementary school, 0 otherwise | 0.51 | 0.39 | 0 | 1 |
| *Female teacher in fifth year.* | Equals 1 if class teacher is female at the fifth grade in elementary school, 0 otherwise | 0.40 | 0.49 | 0 | 1 |
| *Female teacher in sixth year.* | Equals 1 if class teacher is female at the sixth grade in elementary school, 0 otherwise | 0.39 | 0.49 | 0 | 1 |
| *Female teacher 2-6 grades* | Total years of female teacher class between second and sixth grades. | 2.60 | 1.38 | 0 | 5 |
| *Female teacher rate* | The rates of female teachers in respondents' residential areas when they were in the elementary school. | 0.56 | 0.10 | 0.24 | 0.73 |
| *Breakfast childhood* | Number of female candidate in respondents' election district in the 2016 election. | 6.56 | 1.48 | 0 | 7 |
| *Number of children* | Respondent's schooling years | 0.98 | 1.09 | 0 | 6 |
| *Female* | Equals 1 if the respondent is a female, 0 otherwise | 0.50 | 0.50 | 0 | 1 |

Note: Sample is used for estimations in Tables 3 and 4.

Table 2. Mean difference test of quantity of cigarette consumption between male and female teachers homeroom class in each grade.

| | (1) Female teacher | (2) Male teacher | Absolute values |
|---|---|---|---|
| First year. | 1.77 | 2.35 | 3.03*** |
| Second year. | 1.82 | 2.05 | 1.39 |
| Third year. | 1.92 | 1.83 | 0.61 |
| Fourth year. | 1.97 | 1.79 | 1.16 |
| Fifth year. | 2.00 | 1.80 | 1.34 |
| Sixth year. | 1.88 | 1.89 | 0.06 |

Notes: *** denotes statistical significance at the 1 % level

**Table 3**. Regression estimation (Two-limit Tobit): Dependent variable: *Smoke*

|  | (1) | (2) |
|---|---|---|
| *Female teacher in first year.* | −3.24*** <br> (−2.74) | −3.69*** <br> (−3.57) |
| *Female teacher in second year.* | −0.61 <br> (−0.57) |  |
| *Female teacher in third year.* | 0.29 <br> (0.32) |  |
| *Female teacher in fourth year.* | 0.83 <br> (0.89) |  |
| *Female teacher in fifth year.* | 1.00 <br> (0.88) |  |
| *Female teacher in sixth year.* | −0.89 <br> (−0.78) |  |
| *Female teacher 2-6 grades* |  | 0.22 <br> (0.70) |
| *Female teacher rate* | −10.2 <br> (−0.69) | −10.5 <br> (−0.72) |
| *Breakfast childhood* | −0.73*** <br> (−2.67) | −0.72*** <br> (−2.65) |
| *Number of children* | −1.20** <br> (−2.24) | −1.19** <br> (−2.24) |
| *Female* | −11.9*** <br> (−10.3) | −11.9*** <br> (−10.3) |
| Pseudo R-square <br> Left-censored obs <br> Right-censored obs <br> Obs | 0.05 <br> 3966 <br> 6 <br> 4775 | 0.05 <br> 3966 <br> 6 <br> 4775 |

Note: Numbers in parentheses are t-values calculated using robust standard errors. **, and *** indicate significance at the 5 %, and 1 % levels, respectively. Numbers without parentheses are coefficients of each variable. Various control variables are included: Respondent's age and its square term, household income, marital status, job status dummies, educational background dummies, dummies for teaching practices, number of brothers, number of sisters, residential dummies.

**Table 4**. Heckman- model estimation: Dependent variable is *Smoke dummy* in the first and Ln (*Smoke)* in the second stages.

|  | (1) | (2) |
|---|---|---|
|  | First stage |  |
| *Female teacher in first year.* | −0.04*** (−2.94) | −0.05*** (−3.86) |
| *Female teacher in second year.* | −0.01 (−0.90) |  |
| *Female teacher in third year.* | 0.002 (0.20) |  |
| *Female teacher in fourth year.* | 0.01 (0.82) |  |
| *Female teacher in fifth year.* | 0.01 (0.48) |  |
| *Female teacher in sixth year.* | −0.01 (−0.72) |  |
| *Female teacher 2-6 grades* |  | 0.001 (0.17) |
| *Female teacher rate* | −0.11 (−0.64) | −0.11 (−0.65) |
| *Breakfast childhood* | −0.007** (−2.21) | −0.007** (−2.09) |
| *Number of children* | −0.01* (−1.79) | −0.01* (−1.78) |
| *Female* | −0.13*** (−9.69) | −0.13*** (−9.71) |
|  | Second stage |  |
| *Female teacher in first year.* | 0.39 (1.33) | 0.48 (1.47) |
| *Female teacher in second year.* | 0.19 (1.10) |  |
| *Female teacher in third year.* | −0.04 (−0.27) |  |
| *Female teacher in fourth year.* | 0.06 (0.38) |  |
| *Female teacher in fifth year.* | 0.08 (0.45) |  |
| *Female teacher in sixth year.* | 0.003 (0.02) |  |
| *Female teacher 2-6 grades* |  | 0.05 (1.00) |
| *Female teacher rate* | −0.77 (−0.35) | −0.88 (−0.40) |
| *Breakfast childhood* | −0.03 (−0.49) | −0.02 (−0.43) |
| *Female* | 0.33 (0.43) | 0.38 (0.48) |
| Wald Chisquare in the Second Stage | 139.7 | 135.4 |
| First Stage Obs | 4775 | 4775 |
| Second Stage Obs | 809 | 809 |

Note: Numbers without parentheses indicate marginal effects are reported. Numbers in parentheses are z-values. The Z-values are calculated using delta-method standard errors in the first stage. *, **, and *** indicate significance at the 10%, 5 %, and 1 % levels, respectively. Numbers without parentheses are coefficients of each variable. Various control variables are included: Respondent's age and its square term, household income, marital status, job status dummies, educational background dummies, dummies for teaching practices, number of brothers, number of sisters, residential dummies.